\def\lsim{\mathrel{\rlap{\lower4pt\hbox{\hskip1pt$\sim$}}
     \raise1pt\hbox{$<$}}}         
 \def\gsim{\mathrel{\rlap{\lower4pt\hbox{\hskip1pt$\sim$}}
     \raise1pt\hbox{$>$}}}         
\def\na{-\kern-.4em\raise.8ex\hbox{{\tt \scriptsize a}}\ }

\font\fthreei=cmti10 scaled\magstep1
\font\leg=cmssq8

\documentclass[11pt]{article}

\usepackage{amsmath}
\usepackage{amssymb}
\usepackage{graphicx,color}

\begin{document}

\centerline {\Large\bf The Cosmic Microwave Background Spectrum and}

\centerline {\Large\bf an Upper Limit for Fractal Space Dimensionality}

\vspace{1.cm} \centerline{\fthreei F. Caruso~$^{1,2}$ \& V. Oguri~$^{2}$ }
\bigskip
\centerline {$^1$ {Centro Brasileiro de Pesquisas F\'{\i}sicas}}
\centerline {Rua Dr. Xavier Sigaud 150, 22290-180, Rio de Janeiro, RJ, Brazil}
\medskip
\centerline {$^2$ {Instituto de F\'{\i}sica Armando Dias Tavares da Universidade do Estado do
Rio de Janeiro}}
\centerline {Rua S\~ao Francisco Xavier 524, 20550-013, Rio de
Janeiro, RJ, Brazil}
\vspace{0.6cm}
\centerline{emails: francisco.caruso@gmail.com -- oguri@uerj.br}
\vspace{1.5cm}

\abstract{The possibility to constrain fractal space dimensionality from Astrophysics
and other areas is briefly reviewed. Assuming such dimensionality to be $3 +
\epsilon$, a limit to $\epsilon$ can be inferred from COBE satellite data.
The available data for the cosmic microwave background radiation spectrum
are fitted by a Planck's radiation distribution generalized to non integer
space dimensionality. Our analysis shows that the shape of the CMBR
spectrum, which does not depend on the absolute normalization, is correctly
described from this distribution provided the absolute temperature is equal
to 2.726 $\pm$ $0.003\times 10^{-2}$~K and $\epsilon = - (0.957 \pm 0.006)
\times 10^{-5}$. This value for $\epsilon$ is shown to be consistent with
what is found on a very different spatial scale based on a quantum field
phenomenon. The  $|\epsilon|$ is interpreted as an upper limit for how much
space dimensionality could have deviated from three. In other words, this is
the maximum fluctuation space dimensionality should have experienced in a
spatial and temporal scale compared to that of the decoupling era.}

\noindent {\it Subject headings:} cosmic microwave background

\vspace{1.5cm}
\section{Introduction}

The general idea that Astrophysics has something to say about space dimensionality can be traced
back to Kant's conjecture (Handyside 1929) that three-dimensionality of space can be related to the
structure of Newtonian gravitational law. As a second step, one can quote the work of William Paley
(1802), which can be considered the first attempt to shed light on the space dimensionality problem
clearly from Anthropic arguments. In his work, Paley analyzes the consequences of changes in the
form of Newton's gravitational law and of the stability of the solar system on human existence. His
speculations take into account a number of mathematical arguments for an anthropocentric design of
the World, which rest all upon the stability of the planetary orbits in our solar system and on a
Newtonian mechanical {\it Weltanschauung}, as should be expected at that time.

A scientific approach to this issue was introduced by Ehrenfest (1917, 1920) who tried to
understand the threefold nature of space by formulating the question: ``{\it How does it become
manifest in the fundamental laws of Physics that space has 3 dimensions?}". Following this
approach, the first investigations concerning General Relativity and a heuristic model of a
pulsating Universe were made by Tangherlini (1963, 1986), who tried to impose constraints to
integer space dimensionality by searching for bound stable states of the Universe.

The above question can obviously be reversed and we can try to answer the following: ``{\it How do
the fundamental laws of Physics entail space dimensionality?}" (Caruso \& Moreira, 1997). In this
alternative approach, space dimensionality may be taken as an unknown quantity or can be admitted
to have a non integer value $d = 3 + \epsilon$, $\epsilon$ being a parameter to be experimentally
determined. Since the introduction of the concept of fractal dimension by Mandelbrot (1977) this
became an interesting possibility to be explored.

\section{The first predictions}\label{first_prediction}

Following the general aforementioned idea, with just one exception, several authors have determined
limits for $|\epsilon|$, as shown in Table~\ref{Table_limits}.

\begin{table}[htbp]
\caption{\leg First predictions for
deviations $|\epsilon|$ from the integer value 3 for space dimensionality.}
\centerline{\includegraphics[width=12cm]{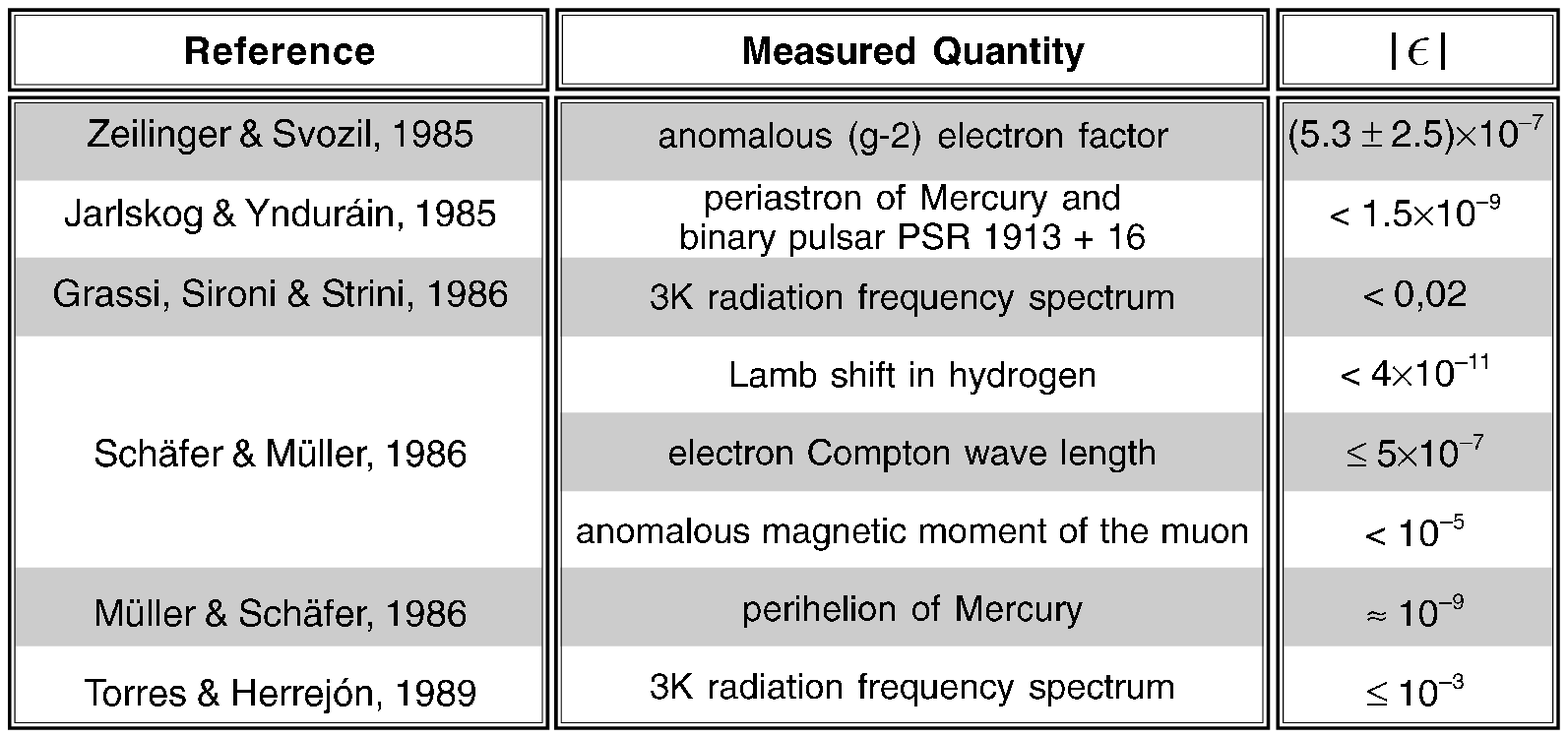}}  \label{Table_limits}
\end{table}

From a general point of view, it is important to stress that all these results cover a very large
length scale from micro to macro cosmos, suggesting that space dimensionality did not fluctuated
significatively from the central value $d=3$ for a very large range of spatial scales. Except form
the two almost similar constraint concerning the cosmic background radiation, the upper limits for
$|\epsilon|$ vary four order of magnitudes for a typical length scale going from $10^{-8}$ (atomic
scale) to $10^{20}$ meters.

So far as astronomical arguments are used, we see from Table~\ref{Table_limits} that different data
from the motion of Mercury gave results of the same order of magnitude, namely, $\epsilon \simeq
10^{-9}$. A quick inspection of this table shows that the two upper limits on $|\epsilon|$ which
follows from analysis of the cosmic background radiation are not so stringent as the others. The
result $|\epsilon|<0.02$ (Grassi, Sironi \& Strini, 1986) was obtained by comparing the available
data at that time for the Relic Radiation against a laboratory black-body source. The limit
$|\epsilon|\leq 10^{-3}$ (Torres \& Herrej\'on, 1989) was achieved estimating the cosmic background
radiation experimental errors for both the radiance and the spectrum to be of the order of ten
percent (Woody \& Richards, 1981; Smoot et al., 1983; Meyer \& Jura, 1984; Uson \& Wilkinson,
1984). The general idea underlining these two papers is that precise measurements of the {\it shape} of the black-body radiation frequency spectrum can be used to set upper limits to deviations of space dimensionality from the integer value 3. In this direction, it was shown by Mather et al. (1990) that the deviation of the shape of the cosmic microwave background spectrum (CMBR) from that of a black-body can be determined with an accuracy much greater than the measurement of the absolute temperature of the sky. In addition, it was concluded, in this paper, that ``{\it since the data show a good null both when FIRAS is loking at the external calibrator and at the sky, one can determine from the interferograms alone that the spectrum of the sky is close to a blackbody, regardless of the details of the data reduction and calibration}". Therefore, we should investigate how the more precise 1996 COBE -- COsmic Background Explorer --  results (Fixsen et al. 1996) can improve the aforementioned limits.

\section{The fit of COBE data}\label{fit}


The FIRAS (Far InfraRed Absolute Spectrophotometer) on COBE satellite has been designed to measure extremely small deviations of the CMBR from a black-body spectrum. However, as all previous experimental results on the frequency spectrum of the Relic Radiation, the COBE data do not provide us with absolute measurements.

This experimental restriction is a consequence of the fact that, at the present moment,  the anisotropy of the CMBR is not large enough to provide us with a clear choice of some particular dark region of the sky which could help us with the calibration (Richards, 1982). However, there is a hope that, at least for the present generation of spectral measurement apparatus, the WMAP (Wilkinson Microwave Anisotropy Probe) data, which provide an accurate enough measurement of the dipole spectrum, could lead to a recalibration of the absolute measurement of the FIRAS COBE using this data (D.J. Fixsen, 2008).


Nevertheless, they led to the conclusion that the CMBR agrees with a black-body spectrum to a high accuracy. The published results (Fixsen et al, 1996) are obtained by a comparison against an almost ideal (emissivity $\approx$~0.98) black-body source placed inside the satellite and calibrated to a second external black-body (emissivity $\approx$~0.99997). Actually, FIRAS, which is a rapid-scan polarized Michelson interferometer, provided a continuous differential comparison with a reference black-body adjusted to null the input signal. Unfortunately, there is no way to measure fluxes independently. From the theoretical point of view, this comparison is accomplished by assuming the Planck distribution and presupposing space to be three-dimensional.

For a $d$-dimensional space, Planck's radiation law for the spectral density energy $u_\nu$, as a
function of temperature $T$ and frequency $\nu$, generalizes to (Caruso \& Oguri, 2006)

\begin{equation}\label{planck_d}
u_\nu = \displaystyle \frac{2(d-1)\pi^{d/2}}{\Gamma(d/2)} \left(\frac{\nu}{c}\right)^d\
\frac{h}{e^{h\nu/kT} - 1}
\end{equation}

For the moment, let us attribute any deviation from the ideal three-dimensional black-body radiation law to be due to the hypothesis that the number of dimension is actually $d=3+\epsilon$ instead of just 3.

\begin{figure}[htbp]
\begin{center}
\begin{minipage}{14.0cm}
\centerline{\includegraphics[width=10cm]{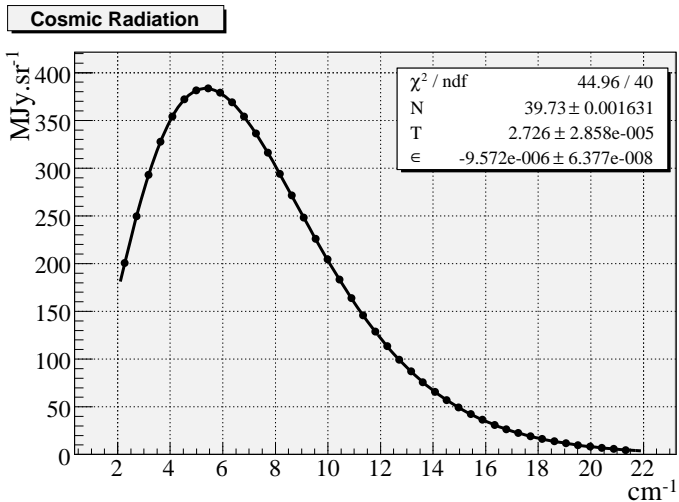}}
\caption{
 \leg Fit of the intensity of cosmic microwave background spectrum as a  function of the inverse  of wavelength by Planck's radiation law
generalized to $d$ dimensions.
}
\end{minipage}
\end{center}
\label{Planck}
\end{figure}

Thus, the data from Fixsen et al. (1996), complemented with those given at
the site of COBE Collaboration (2008), are fitted, using the MINUIT package running the CERN ROOT program, not directly by eq.~(\ref{planck_d}) but by a function of three parameters ($N,\, T,\, \epsilon$)

\begin{equation}\label{fit_function}
u_\nu = \displaystyle N \frac{\nu^d}{e^{h\nu/kT} - 1}
\end{equation}
since what is relevant to our analysis is the shape of the spectrum. The result of the fit is shown in
Figure~\ref{Planck}.

The values which emerge from the fit are:

\begin{equation}
\left\{ \begin{array}{l} \epsilon = - (0.957 \pm 0.006) \times 10^{-5}\ \\
\ \\
T = 2.726 \pm 0.003\times 10^{-2}~\mbox{K}
 \end{array} \right.\label{epsilon}
\end{equation}

Notice that the FIRAS experimental result for the absolute temperature of CMBR, 2.728 $\pm$ 0.004~K
(95\% confidence level), found in Fixsen et al. (1996), is dominated by systematics errors. It is
important to stress that in our fit we have one more parameter besides a slightly different
normalization, which, in practice, did not change the quality of the fit as can be shown by
comparing our formal $\chi^2$ per dof = 1.124 to the 1.15 value found in the COBE analysis.
Actually, the statistical error to be compared to our result for the absolute temperature is
0.00001, given in Table~2 of Fixsen et al. (1996). By other side, the FIRAS-measured CMBR residuals
strongly support the quality of their data and their fit. Thus, it follows that also our result is
statistically significant.

Clearly, the assumption that all deviations from Planck's radiation law might be due to its dependence on dimensionality gives an overestimated value for $\epsilon$. Other sources for these deviations indeed exist and were analyzed Fixsen et al. (1996), such as Bose-Einstein and Compton distortions. In both cases the upper limits found for the involved parameters indicate a small effect. Based on these results, Wright et al. (1994) estimated that from the Bose-Einstein period ($10^5 < z < 3 \times 10^6$) the ratio between the amount of energy converted from anything but the cosmic background to that of the background itself is $\lesssim 6.4 \times 10^{-5}$; for the energy released after that period but before the decoupling era ($z\simeq 10^3$) the limit is quite similar, $\lesssim 6.4 \times 10^{-5}$. Therefore, the value obtained for $\epsilon$ should be seen also as an upper limit, namely $|\epsilon| < 0.957\times 10^{-5}$, related to the epoch when the universe became entirely transparent: the decoupling era.

\section{Discussions}\label{disc}

Strictly speaking, since absolute measurements for the fluxes are not yet available, the physical meaning of the parameter $\epsilon$ should be reinterpreted. Indeed, since the experimental device compares the background radiation to that of a local black-body with controlled temperature, at least in principle, one can figure out that the value of $\epsilon$ could be locally (in this case, where the reference black-body is placed) different from the fractal dimensions of space on the horizon scales. In this case, $\epsilon$ should be interpreted as a difference between these two fractal dimensions at two far away spatial scales, as did Grassi, Sironi \& Strini (1986). Alternatively, one can suppose that locally space is just three-dimensional and, in this case, we are putting a limit on how much space dimensionality could be different from three in the farthest corner of the universe one can look into.

In this paper, we established a bound for $\epsilon$ by fitting data on CMBR taking the absolute temperature $T$, $\epsilon$, and the normalization as free
parameters in eq.~(\ref{fit_function}). The overestimated value we get for
$\epsilon$, equation~(\ref{epsilon}), is two order of magnitude less than the more accurate upper
limit value estimated by Torres \& Herrej\'on (1989) with small errors. Another important difference
found here, in respect to this last paper and to that of Grassi, Sironi \& Strini (1986), is that
we are able to fix also the {\it sign} of the $\epsilon$ parameter, showing that it is {\it
negative}. To the best of our knowledge, this is the first time that the sign of $\epsilon$ can be
determined by analyzing experimental results at astrophysical or cosmological scale. Up to now, the
strongest constraint comes from microphysics. Indeed, Zeilinger \& Svozil (1985) were able to
determine not only the value of $|\epsilon| = |d-3|$ but the sign of this parameter too, from
quantum field theory, getting $ \epsilon = - (5.3 \pm 2.5) \times 10^{-7}$, which is approximately
one order of magnitude less than our upper limit for $|\epsilon|$. It is important to remember that they pointed out that
a value of space-time dimension less than 4 (assuming just one time dimension) would render all the
logarithmic divergences in relativistic quantum field theory finite. Their result seems to resolve
the discrepancies between the theoretical and experimental values of the anomalous magnetic moment
of the electron ({\it cf.} also Svozil \& Zeilinger, 1986). Another evidence in favor of
arbitrarily small non-zero $|\epsilon|$ comes from the study of Ising gauge theories in non-integer
dimensions (Bhanot \& Salvador, 1986). This was a first evidence, based on a numerical analysis,
that for abelian gauge theories, the space-time dimension $d+1 =4$ may, in some sence, be
identified as an upper critical dimension.

In the concluding remarks of their 1985 paper, Zeilinger \& Svozil wrote: ``{\it It is certainly a
challenge for future research to investigate whether or not the deviation of the dimension of
space-time from four can be made more statistically significant than the present work suggests.
Furthermore, the question of possible evidence for such a small deviation in other areas of Physics
deserves attention}". If we assume time to be one-dimensional, as usually done, the main result of
the present paper can be seen as an answer to both challenges.

Finally, since the small deviation from three-dimensionality of space obtained here is extract
directly from the CMBR data, it actually suggests that space dimensionality did not vary significantly in a
huge temporal scale, once this background radiation is expected to be related to the Big Bang. This time scale can be safely put on the later epoch where the universe was about $3\times 10^{5}$ yr old ($z \simeq 10^{3}$). In
addition, due to its isotropy, this is the {\it only} experimental situation that, being confronted
to {\it any} other local experiment on Earth, aimed to measure the number of space dimensions,
drive to the conclusion that dimensionality did not vary also in a very large spatial scale (a
cosmological scale).


\section{References}\label{ref}

\paragraph*{}

\begin{enumerate}

\item Bhanot, G. \& Salvador, R. 1986, Phys. Lett. 167B, 343.

\item Caruso, F. \& Moreira Xavier, R. 1997, in Scorzelli, R.B.,
Azevedo,  I.S. \& Baggio Saitivitch, E. (eds.), Essays on Interdisciplinary Topics in Natural Sciences
Memorabilia: Jacques A. Danon. (Gif-sur-Yvette/Singapore: Editions Fronti\`eres).

\item Caruso, F. \& Moreira Xavier, R. 1987, Fundamenta Scienti\ae\ 8, 73.

\item Caruso, F. \& Oguri, V. 2006, Modern Physics (in Portuguese), (Rio de Janeiro: Elsevier).

\item COBE Collab., http://lambda.gsfc.nasa.gov/product/cobe/firas\underline{\ }monopole\underline{\ }get.cfm,
accessed April, 2008.

\item Ehrenfest, P. 1917, Proc. Amsterdam Acad. 20, 200 (1917); reprinted in M.J. Klein (ed.),
Paul Ehrenfest -- Collected Scientific Papers. (Amsterdam: North Holland), pp. 400-409 (1959).

\item --------- 1920, Annalen der Physik 61, 440.

\item Fixsen, D.J. et al. 1996, Ap. J. 473, 576.

\item Fixsen, D.J. 2008, private communication.

\item Grassi, A., Sironi, G. \& Strini, G. 1986, Astrophys. and Space Science 124, 203.

\item Handyside, J., 1929, Kant's inaugural dissertation and the early writings on space.
 (Chicago: Open Court), reprinted by Hyperion Press, 1979.

\item Jarlskog, C. \& Yndur\'ain, F.J. 1986, Europhys. Lett. 1, 51.

\item Mandelbrot, B.B. 1977, The Fractal Geometry of Nature (New York: Freeman and Co.).

\item Mather, J.C. et al. 1990, Ap. J. (Letters) 354, L37.

\item Meyer, D.M. \& Jura, M. 1984, Ap. J. (Letters) 276, L1.

\item M\"uller, B. \& Sch\"afer, A. 1986, Phys. Rev. Lett. 56, 1215.

\item Paley, W. 1802, Natural Theology. A recent book's reprint is available from Oxford University Press, 2006.

\item Richards, P.L. 1982, Pil. Trans. R. SOc. Lond. A307, 77.

\item Sch\"afer, A. \& M\"uller, B. 1986, J. Phys. A: Math. Gen. 19, 3891.

\item Smoot, G.F. et al. 1983, Phys. Rev. Lett. 51, 1099.

\item Svozil, K. \& Zeilinger, A. 1986, Int. J. Mod. Phys. A 1, 971.

\item Tangherlini, F.R. 1963,Nuovo Cimento 27, 636.

\item --------- 1968, Nuovo Cimento 91B, 209.

\item Torres, J.L. \& Herrej\'on, P.F. 1989, Revista Mexicana de F\'{\i}sica 35, 97.

\item Uson, J.M. \& Wilkinson, D.T. 1984, Ap. J. (Letters) 277, L1.

\item Woody, D.P. \& Richards, P.L. 1981, Ap. J. 248, 18.

\item Wright, E.L. et al. 1994, ApJ, 420, 450.

\item Zeilinger, A. \& Svozil, K. 1985, Phys. Rev. Lett. 54, 2553.
\end{enumerate}

\end{document}